\documentclass[twocolumn,twoside,slac]{revtex4}
\usepackage{graphicx}
\usepackage{fancyhdr}
\begin{document}

\title{Data compression using correlations and stochastic processes in the ALICE Time Projection chamber}

\author{M.Ivanov}
\affiliation{CERN, Switzerland}
\author{A.Nicolaucig}
\affiliation{Ecole Polytechnique Federale Lausanne, Switzerland}
\author{A.Krechtchouk}
\affiliation{Lomonosov Moscow State University, Russia }

\begin{abstract}

In this paper lossless and  a quasi lossless algorithms for the
online compression of the data generated by the Time Projection
Chamber (TPC) detector of the ALICE experiment at CERN are
described.

The first algorithm is based on a lossless source code modelling
technique, i.e. the original TPC signal information can be
reconstructed without errors at the decompression stage. The
source model exploits the temporal correlation that is present in
the TPC data to reduce the entropy of the source.

The second algorithm is based on a lossy source code modelling
technique, i.e. it is lossy if samples of the TPC signal are
considered one by one. Nevertheless, the source model is
quasi-lossless from the point of view of some physical quantities
that are of main interest for the experiment. These quantities are
the shape, the location of the center of gravity as well as the
total charge of the signal.

In order to evaluate the consequences of the error introduced by
the lossy compression, the results of the trajectory tracking
algorithms that process data offline are analyzed, in particular,
with respect to the noise introduced by the compression. The
offline analysis has two steps: cluster finder and track finder.
The results on how these algorithms are affected by the lossy
compression are reported.

In both compression technique entropy coding is applied to the set
of events defined by the source model to reduce the bit rate to
the corresponding source entropy. Using TPC simulated data, the
lossless and the lossy compression achieve a data reduction to
49.2\% of the original data rate and respectively in the range of
35\% down to 30\%  depending on the desired precision.

In this study we have focused on methods which are easy to
implement in the frontend TPC electronics.

\end{abstract}

\maketitle

\thispagestyle{fancy}


\section{Introduction}

ALICE (A Large Ion Collider Experiment) is an experiment that will
start in 2007 at the LHC (Large Hadron Collider) at CERN
\cite{TPCTDR,alicewww}.  The experiment will study collisions
between heavy ions with energies around 5.5 TeV per nucleon.  The
collisions will take place at the center of a set of several
detectors, which are designed to track and identify the produced
particles.

\begin{figure*}[t]
\centering
\includegraphics[width=70mm,angle=-90]{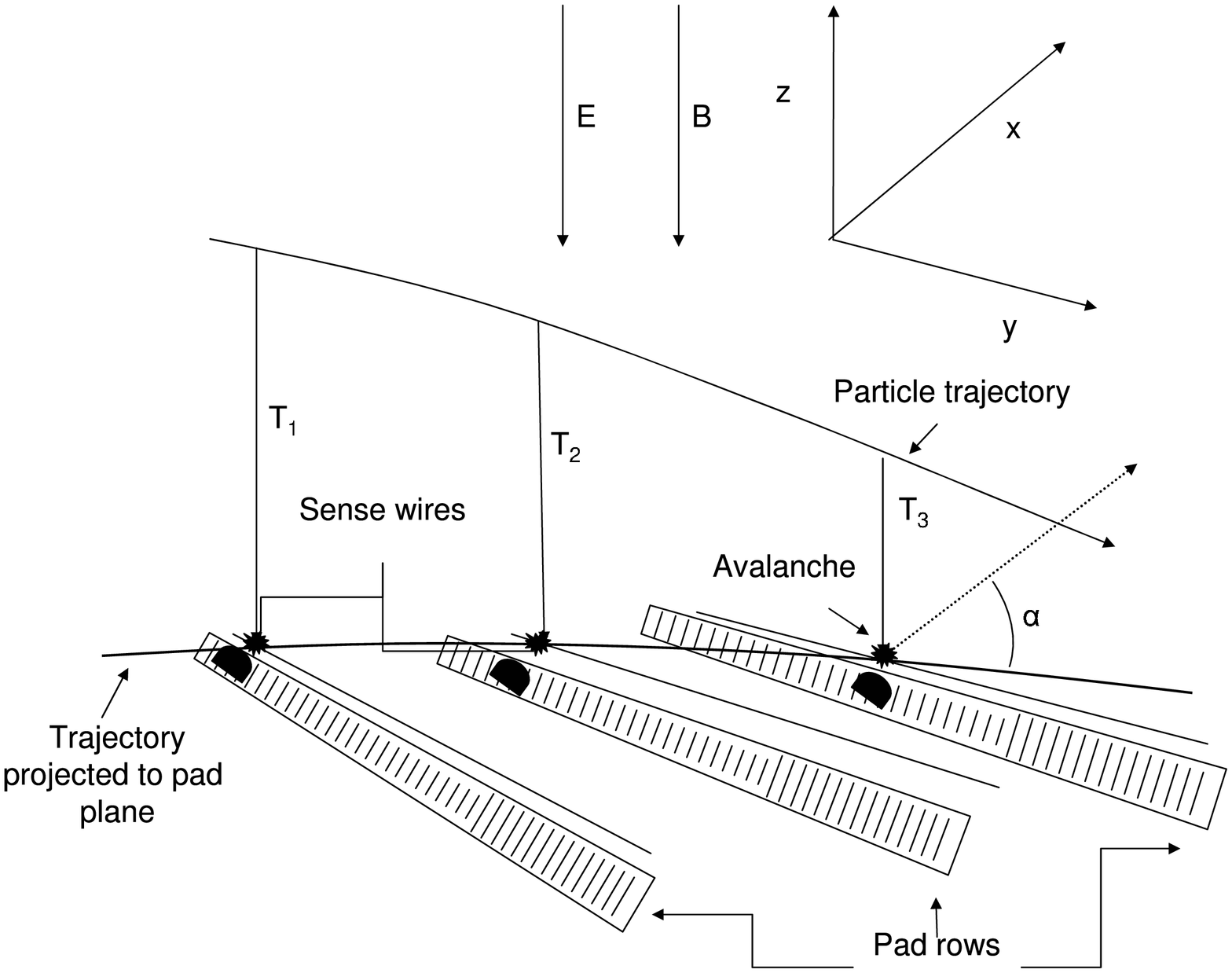}
\includegraphics[width=70mm,angle=-90]{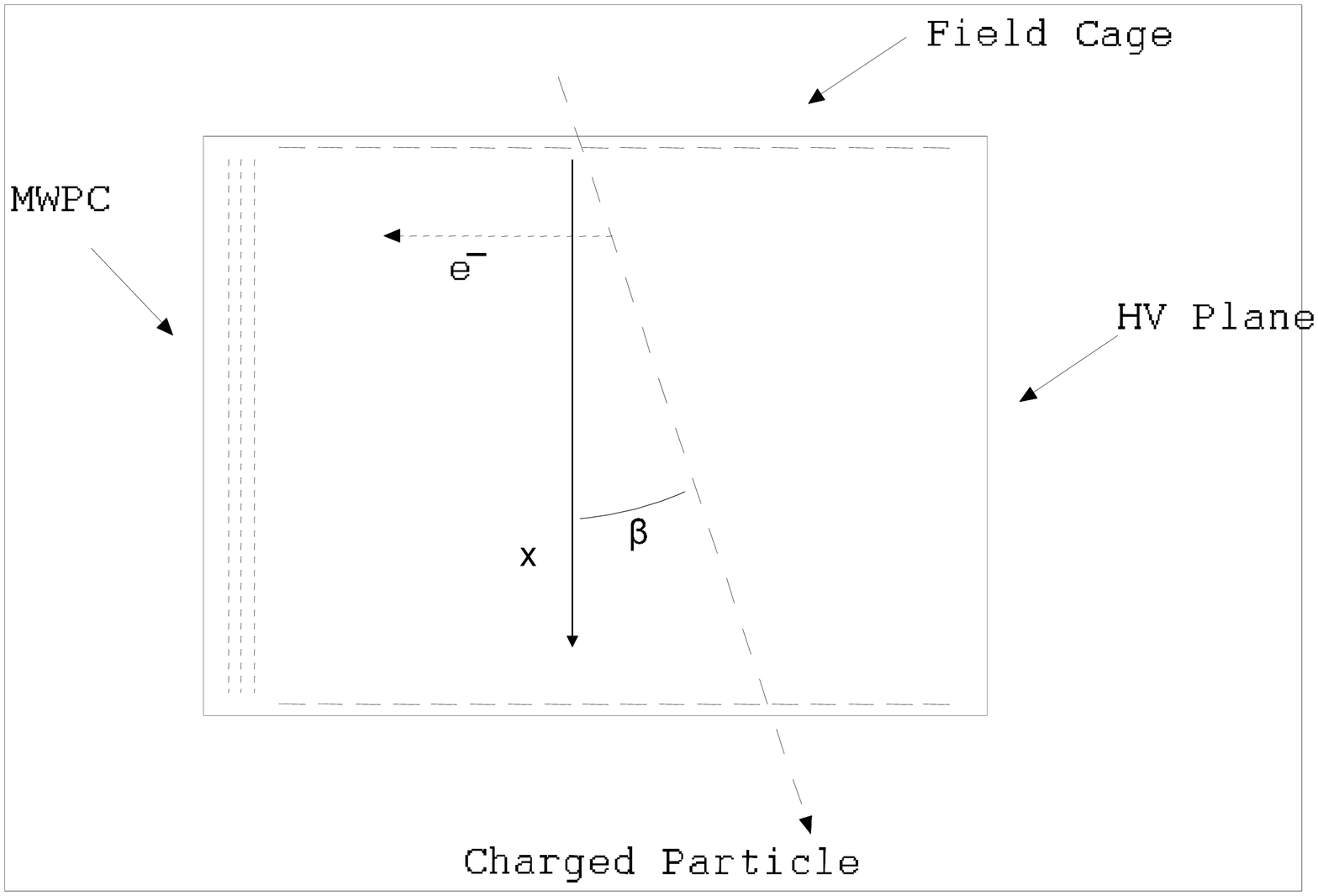}
\caption{Schematic view of the detection process in TPC  (upper
part - perspective view, lower part - side view).} \label{figTPC}
\end{figure*}

One of the main detectors of the ALICE experiment is the Time
Projection Chamber (TPC). Its task is track finding, momentum
measurement and particle identification by d$E$/d$x$. Good
two-track resolution, required for correlation studies, is one of
the main design goals.

The TPC is a large horizontal cylinder, filled with gas, where a
suitable axial electric field is present. When particles pass
through, they ionize the gas atoms, and the resulting electrons
drift in the electric field. By measuring the arrival of electrons
at the end of the chamber, the TPC can reconstruct the path of the
original charged particles. The electrons are collected by more
than 570~000 sensitive pads where they create signals. These
signals are amplified by a preamplifier--shaper and digitalized by
a 10-bit A/D converter at a sampling frequency of 5.66~MHz.  The
digitalized signal is processed and formatted by an Application
Specific Integrated Circuit (ASIC) called ALTRO (ALICE TPC
Read-Out) \cite{musa00altro}.  At this stage, the overall
throughput of the 570~000 channels is around 8.4~GByte/s.

The total amount of the TPC data is expected to be about 1 PBy per
year. In order to keep the complexity and cost of the data storage
equipment as low as possible, we have to reduce the volume of data
using suitable data compression methods. The cost reduction of the
data storage system is roughly proportional to the data
compression factor. Furthermore, it is better to implement the
compression system in the front-end electronics at the output of
the ALTRO circuit, so that the cost for the optical links, which
carry data out of the chamber to the following stages of the
acquisition chain, could be also reduced.More sophisticated
methods for TPC data compression based on online tracking, which
will be used further in data acquisition chain are developed in
Bergen and Heidelberg \cite{bergen}.

The use of a lossy source model, justified by the fact that
generally it can provide significantly higher compression ratios
compared to lossless models, has the drawback that some deterioration in the
reconstruction of data must be accepted. Lossy source models have
become very popular in the last decade in the field of audio and
video compression for their remarkable performance.
Lossy models have been carefully designed so that reconstruction
distortions are not perceived using psychovisual or psychoacoustic
models or  they remain comparable with the intrinsic signal noise.

Obviously, for physical data, psychovisual or psychoacoustic tests
are meaningless or even not applicable since the TPC signal is not
to be observed by the human eye or ear. In \cite{nicolaucig02},
the compression noise introduced on the sample values by the
described lossy or quasi lossless techniques has been evaluated in
terms of RMS of the introduced Error (RMSE).

However, in this case, the RMSE, despite being a simple and well
known distortion measure, is not very useful. The fundamental
information that has to be extracted from TPC data are not sample
values themselves but the physical quantities that enable the
reconstruction of particle trajectories. Therefore,  the correct
way to evaluate the importance of the distortion introduced by the
compression--decompression process has to be related to the high
level information that is carried by the data. In particular, TPC
data are collected with the objective of measuring particle energy
and trajectory.

Therefore, the most effective way to estimate the consequences of
the compression distortion error, is to observe how the extraction
of energy and trajectories are affected by the
compression--decompression process. A simple way to obtain these
estimates is to apply the cluster finding and tracking algorithms
on both simulated data and their compressed--decompressed version
and compare the results.

 This article is arranged in the following
way. In section~\ref{sec:Stochastic}, all stochastic processes
relevant for particle detection in ALICE TPC are briefly
described. In section~\ref{sec:Altro}, TPC data format is
specified. In section~\ref{sec:Lossless}, different lossless
compression techniques are described and their efficiencies are
compared. In section~\ref{sec:Lossy}, the fast one dimensional
lossy compression technique is shown and the impact of
compression--decompression to the distortion of most important
physical quantities is demonstrated.

\section{\label{sec:Stochastic}Stochastic processes in TPC}

\subsection{Ionization in gas}

A charged particle that traverses the gas of the chamber leaves a
track of ionization  along its trajectory.  The collisions with
the gas atoms are purely random. They are characterized by a mean
free path $\lambda$ between ionizing encounters, which is given by
the ionization cross-section per electron $\sigma_{\rm{ion}}$ and
the density {\it{N}} of electrons:
\[\lambda=1/(N\sigma_{\rm{ion}}).\]
Therefore, the number of encounters along the length {\it{L}} has
the mean of $L/\lambda$, and the frequency distribution is given
by Poisson distribution
\[P(L/\lambda, k)=\frac{(L/\lambda)^k}{k!}\exp(-L/\lambda).\]
The mean free path $\lambda$ is given by the properties of the gas
and by charged particle characteristics:
\[\lambda = \frac{1}{N_{\rm{prim}}{\times}f(\beta\gamma)},\]
where $N_{\rm{prim}}$ is the number of primary electrons per cm
produced by a Minimum Ionizing Particle (MIP), and
$f(\beta\gamma)$ is Bethe--Bloch curve.

\subsection{Generation of secondary electrons}
The energy loss $E_{\rm{tot}}$ released in primary ionization to
atomic electrons is a random variable. It can be described by
Photo-Absorbtion Ionization model (PAI). In most cases, if one
neglects the atomic shell structure, at sufficiently high
$E_{\rm{tot}}$ (the energy where the atomic shell structure is not
more important) it obeys $1/E_{\rm{tot}}^2$ rule.

If the electron produced by the charged particle has sufficient
kinetic energy $E_{\rm{tot}}$, it will produce secondary electrons
creating thus electron cluster. The mean total number of electrons
in such cluster is given by:
\[N_{\rm{tot}}=\frac{E_{\rm{tot}}-I_{\rm{pot}}}{W_{\rm{ion}}} + 1,\]
where $E_{\rm{tot}}$ is the energy loss in a primary collision,
$W_{\rm{ion}}$ is the effective energy required to produce an
electron--ion pair and $I_{\rm{pot}}$ is the first ionization
potential. The random character of the secondary ionization
process smears out structures in $E_{\rm{tot}}$ spectra, atomic
shell structure behavior is suppressed. For example in the gas
mixture 90\% Ne, 10 \% CO$_2$ the $E^{-2.2}_{\rm{tot}}$ effective
parametrization at lower $E_{\rm{tot}}$ can be used.

\subsection{Diffusion of electrons}
Produced electrons drift through the gas with an effective
constant drift velocity  in the direction given by the electric
field $\mathbf{E}$ and magnetic field $\mathbf{B}$ (which we
assume are parallel to {\it{z}}-direction). Drifting electrons are
scattered on the gas molecules so that their direction of motion
is randomized in each collision. The position of the electron,
after drifting over a distance $L_{\rm{drift}}$, can be described
by 3-D Gaussian distribution:
\begin{eqnarray}
P(x,y,z)=\frac{1}{\sqrt{2\pi}\sigma_{\rm{T}}}\exp\left[-\frac{(x-x_0)^2}{2\sigma_{\rm{T}}^2}\right] \nonumber \\
\frac{1}{\sqrt{2\pi}\sigma_{\rm{T}}}\exp\left[-\frac{(y-y_0)^2}{2\sigma_{\rm{T}}^2}\right] \nonumber\\
\frac{1}{\sqrt{2\pi}\sigma_{\rm{L}}}\exp\left[-\frac{(z-L_{\rm{drift}})^2}{2\sigma_{\rm{L}}^2}\right],
\end{eqnarray}
where $\{x_0, y_0, z_0\}$ is the electron creation point and
transversal diffusion $\sigma_{\rm{T}}$ respectively longitudinal
diffusion $\sigma_{\rm{L}}$ are given by drift length
$L_{\rm{drift}}$ and gas coefficient $D_{\rm{T}}$ and $D_{\rm{L}}$
\[\sigma_{\rm{T}}=D_{\rm{T}}\sqrt{L_{\rm{drift}}},\]
\[\sigma_{\rm{L}}=D_{\rm{L}}\sqrt{L_{\rm{drift}}}.\]

\subsection{$\mathbf{E{\times}B}$ and unisochronity effect near the anode wires}
It has been assumed that the electric and magnetic fields in the
drift volume are uniform and parallel. This, however, is not true
close to the anode wires, where the electric field becomes radial.
Thus the electrons experience a shift along the wire direction
(due to the Lorentz force). If an electron enters the readout
chamber at the point $(x_{\rm{e}},y_{\rm{e}})$, it is displaced in
the {\it{x}}-direction (assuming that the wires are placed along
{\it{y}}-axis). The new {\it{y}}-position of the electron is then
given by
\[ y = y_{\rm{e}} + \omega \tau \cdot (x-x_{\rm{e}}) \: , \]
where $x$ is the coordinate of the wire on which an electron is
collected,
 and $\omega \tau$ is the tangent of Lorentz angle ($\mathbf{E{\times}B}$ effect).
The drift length which determines {\it{z}} coordinate will be also
affected, because of change in the path to the anode wire
(unisochronity effect).

\subsection{Signal generation}
Inside the readout chamber, as an electron drifts towards the
anode wire, it travels in an increasing electric field. Once the
electric field is strong enough that between collisions with the
gas molecules the electron can pick up sufficient energy for
ionization, another electron is created and the avalanche starts.
As the number of electrons multiplies in successive generations,
the avalanche continues to grow until all the electrons are
collected on the wire. The resulting number of electrons created
in the avalanche, can be described by an exponential probability
distribution
\[ P(q) = \frac{1}{\overline{q}} \cdot \exp { - \frac{q}{\overline{q}}} \: , \]
where $\overline{q}$ is the average avalanche amplitude.

An electron avalanche collected on the anode wire induces a charge
on the pad plane. This charge is integrated over the pad area. The
time signal is obtained by folding the pad response to the
avalanche with the shaping function of the
preampamplifier--shaper. This signal is then sampled with a
constant frequency. On the top of sampled signal a random
electronic noise is superimposed.

As a result a charged particle interacting with gas generates a
cluster of amplitudes. This cluster is used for later estimation
of local track position and of local energy deposition. The shape
of the cluster is used as additional information for the
estimation of position uncertainties and for the estimation of the
overlap factor between two tracks.

\subsection{Accuracy of local coordinate measurement}

The accuracy of the coordinate measurement is limited by a track
angle which spreads ionization and by diffusion which amplifies
this spread.

The track direction with respect to pad plane is given by two
angles $\alpha$ and $\beta$ (see fig.~\ref{figTPC}). For the
measurement along the pad-row, the angle $\alpha$ between the
track projected onto the pad plane and pad-row is relevant. For
the measurement of the the drift coordinate ({\it{z}}--direction)
it is the angle $\beta$ between the track and {\it{z}} axis.

The ionization electrons are randomly distributed along the
particle trajectory. Fixing the reference {\it{x}} position of a
electron at the middle of pad-row, the {\it{y}} (resp. {\it{z}})
position of the electron is random variable characterized by
uniform distribution with the width $L_{\rm{a}}$, where
$L_{\rm{a}}$ is given by the pad length $L_{\rm{pad}}$ and the
angle $\alpha$ (resp. $\beta$):
\[L_{\rm{a}}=L_{\rm{pad}}\tan\alpha\]
The diffusion smears out the position of the electron with
gaussian probability distribution with $\sigma_{\rm{D}}$.
Contribution of the $\mathbf{E{\times}B}$ and unisochronity effect
is in the case of Alice TPC negligible.

The accuracy of the position measurement  can be expressed as:

{$\sigma_{\rm{z}}$} of cluster center in z (time) direction:
\begin{eqnarray}
     \lefteqn{\sigma^2_{\rm{z_{COG}}} = \frac{D^2_{\rm{L}}L_{\rm{drift}}}{N_{\rm{ch}}}G_{\rm{g}}+{}}\nonumber\\
        &&{}\frac{\tan^2\alpha~L_{\rm{pad}}^2G_{\rm{Lfactor}}(N_{\rm{prim}})}{12N_{\rm{chprim}}}+
        \sigma^2_{\rm{noise}}
         \label{eq:ResZ1}
\end{eqnarray}

and {$\sigma_{\rm{y}}$} of cluster center in y(pad) direction:
    \begin{eqnarray}
     \lefteqn{\sigma^2_{\rm{y_{COG}}} = \frac{D^2_{\rm{T}}L_{\rm{drift}}}{N_{\rm{ch}}}G_{\rm{g}}+{}} \nonumber\\
        &&{}\frac{\tan^2\beta~L_{\rm{pad}}^2G_{\rm{Lfactor}}(N_{\rm{prim}})}{12N_{\rm{chprim}}}+
        \sigma^2_{\rm{noise}}
        \label{eq:ResY1}
    \end{eqnarray}
 where
${N_{\rm{ch}}}$ is the total number of electrons in cluster,
${N_{\rm{chprim}}}$ is the number of primary electrons in cluster,
${G_{\rm{g}}}$ is the gas gain fluctuation factor parametrization,
${G_{\rm{Lfactor}}}$ is the secondary ionization fluctuation
factor and $\sigma_{\rm{noise}}$ describe the contribution of the
electronic noise and ADC quantization to the resulting sigma of
the COG.

The typical resolution in the case of ALICE TPC is on the level of
$\sigma_{\rm{y}}\sim$~0.8mm and $\sigma_{\rm{z}}\sim$~1.0mm
integrating over all clusters in the TPC.

\subsection{Accuracy of the total amplitude measurement}

The total charge deposited in the clusters can be used for
particle identification. The important value, which is specific
for different particle types and different particle momenta, is
the number of primary collision per unit  length,
$N_{\rm{chprim}}$. $N_{\rm{chprim}}$ is a random variable
described by Poisson distribution. Due to the secondary ionization
and gas gain fluctuations the total charge is described by very
broad Landau distribution.

\section{\label{sec:Altro}The ALICE TPC read-out data format}

Before describing the compression algorithm, it is necessary to
spend a few words on the format of data at the output of ALTRO
circuit, in order to understand how the compression algorithms are
applied. Such data are indeed the input of the compression system
\cite{TPCTDR,musa00altro}.

In the ALTRO data format only the samples over a given threshold
are considered, while the others are discarded. This means that,
if we call {\em bunch} a group of adjacent over-threshold samples
coming from one pad,  the signal can be represented ``bunch by
bunch''.  More precisely, a bunch is described by three fields:
temporal information (temporal position of the last sample in the
bunch), one 10-bit word, bunch length (i.e. the number of samples
in the bunch, one 10-bit word), and sample amplitude values (few
10-bit words).

\section{\label{sec:Lossless}Lossless compression of TPC signals }

The lossless techniques of the data compression are based on the
fact that TPC sample values (ADC and temporal) are not equally
probable.  A  theoretical lower limit on the average word size
using Huffman codding, or arithmetic coding lossless technique  is
given by entropy of the data source:

\begin{eqnarray}\
E(p) = \sum{p(A)\log_{\rm{2}}p(A)}
         \label{eq:Entropy}
\end{eqnarray}
The lossless techniques described in this paper are based mainly
on an appropriate probability model for each data field of the
ALTRO data format. Specific probability models for each sample in
a bunch were developed. These models intend to capture both
temporal correlation among samples and the characteristic shape of
TPC electrical pulses.

\subsection{Time information}

 As already mentioned, in the
ALTRO data format time information is represented as the 10-bit
number of the time-bin of the last sample of the bunch. The
probability distribution of this variable is roughly uniform. In
order to achieve better compression ratio this variable is
substituted by the distance between two consecutive bunches. The
probability of this variable is described by exponential
distribution with much lower entropy factor. The entropy of
temporal information is given by mean distance between two
bunches. It depends on the event multiplicity, noise level and
local occupancy, which is known function of the pad-row radius. In
order to optimize entropy coding, it will be necessary to
investigate probability distribution as a function of track
multiplicity. This information will be known from other faster
ALICE detectors.

The mean number of bits used for the coding of time information is
roughly 4.9 bits for the full event with maximal track density.
Using different codes in different places inside TPC, an
additional 6\% reduction in time information can be achieved.

\subsection{Bunch length}

In the ALTRO data format, the bunch length is represented as a
10-bit code number of samples in the bunch. The bunch length
depends on the diffusion, the angular effect and the total
deposited energy. There is no apparent correlation with data coded
before. Small diffusion for short drift length is compensated by
big angular effect. The total deposited energy is known only after
coding of the bunch length.
 Since no apparent correlation with other data (e.g. length of
adjacent bunches) exists and no better model (i.e. a model of
events with lower entropy) could be found, this information is
coded directly.

\subsection{Sample values coding}

Sample values are the main contribution to the resulting data
volume. This subsection describes, first a basic model, and then
introduces a more sophisticated one, that can provide higher
performances in terms of compression efficiency.

Data compression can be obtained by directly applying entropy
coding to the sample values without any modelling of the
information source. This method will be referred bellow (in
table~\ref{tabLossless}) as  Entropy Coding (EC).

\subsection{Coding model based on the sample position}

Improvements in compression performance can be obtained by
appropriate modelling. A first improvement has been achieved by
the fact that the statistics of the signal sample values depend on
the position of the sample itself in the bunch.

\begin{table}[t]
\begin{center}
\begin{tabular}{|l|c|c|c|c|c|c|}
\hline \textbf{Bunch length} &\textbf{freq} &\textbf{0} &
\textbf{1} & \textbf{2}& \textbf{3} & \textbf{4}\\
 \hline
1 & 136    & 2.21 &&&&\\
\hline
2 & 279    & 4.04 & 4.04 &&&   \\
\hline
3 & 422    & 4.64 & 5.5 & 4.64&&  \\
\hline
4 & 241    & 4.18 & 6.67 & 6.02 & 4.18& \\
\hline
5 & 53     & 3.83 & 6.1  & 7.15 & 6.1 & 3.83\\
\hline
\end{tabular}
\caption{\label{tab:SampleEntropy}Entropy of the sample data as a
function of the sample position in the bunch. Frequency of the
sample length is given in arbitrary units.}
\end{center}
\end{table}

Due to the pseudo Gaussian shape of most of the bunches, the first
and the last sample of each bunch are likely to have a smaller
value with respect to those in central positions. Similarly, small
values are also expected for isolated samples, i.e. belonging to
one-sample bunches (see table \ref{tab:SampleEntropy}).

Therefore, a classification of the samples into three classes was
chosen: one class for isolated samples, one for samples at the
beginning and at the end of multiple sample bunches, and the last
for samples in the central positions of a bunch.  Using three
different probability distributions for entropy coding the sample
values can be coded more efficiently than using only one
probability distribution. This coding scheme will be referred in
table \ref{tabLossless} as  coding using Sample Position (SP).

\subsection{Source models exploiting temporal correlation}
Improvement on compression performances can be expected by
exploiting temporal correlation, i.e. the correlation between
consecutive samples; this can be done by implementing a suitable
prediction scheme.

This  approach is explained on the example, where a three-sample
bunch is considered. Let us assume that the first two samples have
already been coded and that the third one has to be coded. The
code to be used for sample No. 3 may be chosen among eight
possible codes according to the value of sample No. 2. In
particular, this is done by subdividing the range of sample No. 2
(i.e. 0. . . 1023) into different intervals, and associating a
different code (for the third sample) to each of these intervals.

This conditioned probability model can be extended to all the
samples that are not in the first position in the bunch and for
any bunch length. However, if the real-time implementation
constraints are taken into account, and, in particular, the need
to reduce the memory size of the model, it is not good to have an
exceedingly large number of codes. Consequently, samples are
partitioned into four classes only, to keep the complexity of the
model low. This limitation does reduce the efficiency of the model
but the reduction is only of the order of 0.6\%. This coding
scheme will be referred in table \ref{tabLossless} as  coding
using Temporal Correlation (TC).

\subsection{\label{sec:Losslescomp}Comparison of different lossless technique}

\begin{table}[t]
\begin{center}
\begin{tabular}{|l|c|c|c|c|}
\hline \textbf{} & \textbf{Time} & \textbf{Length} &
\textbf{Samples}& \textbf{Total}
\\
\hline
Altro&10 bits&10 bits&38.1 bits& 58.1 bits(100\%) \\
\hline
EC&4.9 bits&3.1 bits&22.4 bits&30.3 bits (52.5\%)\\
\hline
SP&4.9 bits&3.1 bits&21.3 bits&29.2 bits (50.3\%)\\
\hline
TC&4.9 bits&3.1 bits&20.7 bits&28.6 bits (49.2\%)\\
\hline
\end{tabular}
\caption{\label{tabLossless} Performance of several lossless
techniques compared to the zero suppressed ALTRO data format.
ALTRO: original ALTRO data; EC: entropy coding of sample values,
bunch length, and time information; SP: classification of samples
according to their position (3 code tables used); TC: coding
technique that exploits temporal correlation (20 code tables
used). Numbers in the columns represent the number of bits per
bunch dedicated to each field; numbers in the right column
represent the overall number of bits per bunch, and, in
parenthesis, the size with respect to the original ALTRO data
format. }
\end{center}
\end{table}

The results of different lossless approaches on simulated TPC data
are shown in table \ref{tabLossless}. It may be noticed that the
latter TC technique provides a compression of data down to 49.2\%
of the original size. Even this best technique provides reduction
factor only by 3\% better then direct EC technique.

Additional attempt tried to use predicted mean cluster shape
information. Knowing the position of the bunch, the diffusion
given by drift length ($L_{\rm{drift}}$) and inclination angle for
primary particles are known. However, due to the fluctuation of
cluster shape and due to the large amount of secondary particles
with unknown angles, this prediction is not very good, and the
entropy of the samples is reduced only by additional factor 2\%.

\subsection{Space correlation}

In the trial to exploit space correlation, three lossless models
have been considered. The first is based on spatially conditioned
probability, the second on a predictive model, third on
2-dimensional cluster finder, with residual saving.

The first one is the equivalent, in the spatial domain, to what
has been done for time correlation. Different codes are available
to code the samples; for each sample, the appropriate code is
selected according to the value of the samples in the same
time-bin but in adjacent pads. This method provides poorer
performance when compared with the one which exploits time
correlation (the comparison being done using the same model
complexity, i.e. number of probability distributions available in
memory). Moreover, these two techniques cannot be easily combined,
i.e. it is difficult to exploit both temporal and spatial
correlations at the same time, because this would require a very
large number of probability distributions (i.e. code tables).

The second method that has been investigated uses the prediction
of the sample values from the samples in adjacent pads and coding
the error of this prediction. Unfortunately, also for this model,
the performance is not very good.

Pulses in one pad-row often resemble temporally shifted versions
of those in the adjacent pad-row. The two methods described above
have been modified by adding the first stage which shifts pulses
so as to increase spatial correlation with adjacent. Although the
performance has slightly improved, the increase of the compression
efficiency was lower than expected. The correlations are
relatively small. The main problems here are in the big amount of
secondary particles crossing TPC with unknown $\beta$ angle (not
pointing to the primary vertex), big spread of the particle
momenta (unknown $\alpha$ angle) and the Landau fluctuation of
deposited energy on different pad rows, which is almost
uncorrelated. Moreover, the position of the original track
relative to the pad, affects the correlation by a large factor.
The signal amplitude in adjacent pads and adjacent pad-rows are
very weakly correlated, unless the position and direction of the
track is known.

In order to get better knowledge of the track position,
two-dimensional cluster finding can be done before. The entropy of
the stored residuals is by 30\% lower than entropy of the original
samples  but there are problems with the track overlaps and with
description of the cluster topology (i.e. where to store
residuals).

Based on these results we conclude that it is not simple to
exploit spatial correlation (i.e. correlations between adjacent
channels). There might be more sophisticated and complex lossless
models able to exploit it, but relatively simple models seem to
fail.

\section{\label{sec:Lossy}Lossy compression of TPC signals }

\subsection{Fluctuation and accuracy of the amplitude measurement}

The number of primary ionization electrons produced by the charged
particle in the gas is the random variable described by Poisson
distribution with the mean value ~14.35 $\rm{cm^{-1}}$  for
minimum ionizing particle  in the gas of Alice TPC. The secondary
electron production (described by ~$E^{-2.2}$ probability
distribution) increases the number of produced electrons. Maximum
probable value is  25 $\rm{cm^{-1}}$ of total electrons. This
effect also smears probability distribution to the relatively
broader Landau distribution.

Due to the angular effect and diffusion, electrons are distributed
among several time-bins and pads. The number of electrons which
contributes to the given pad and time-bin is described roughly by
Poisson distribution. Each of the registered electrons is subject
of gas multiplication which is described by exponential
probability distribution. Over this, additional electronic noise
is superimposed to each signal.

If we fix the track position and the number of primary electrons,
the remaining sample uncertainties can be in the first
approximation estimated as:
\begin{eqnarray}\
     \sigma_{\rm{S}} = \sqrt{\sigma_{\rm{noise}}^2+G{\times}A}
        \label{eq:Sample}
\end{eqnarray}
where $\sigma_{\rm{noise}}$ is given by electronic noise
 and sampling imprecision, and {\it{G}} is the gain
conversion factor.

The situation is more complicated, data samples are correlated
through the time response function and the pad response function.
The relative correlation between the samples depends on the ratio
of the width of the response functions to the width given by
stochastic processes.

\begin{table}[t]
\begin{center}
\begin{tabular}{|l|c|c|c|c|}
\hline \textbf{} & \textbf{no} & \textbf{$K_{\rm{off}}$=1}
&\textbf{$K_{\rm{off}}$=1.5}&\textbf{$K_{\rm{off}}$=1}\\
\textbf{} & \textbf{} & \textbf{$K_{\rm{cor}}$=1}
&\textbf{$K_{\rm{cor}}$=1.5}&\textbf{$K_{\rm{cor}}$=2}\\
\hline
range&0..1024&0..62&0..42&0..33\\
\hline
entropy&5.7&3.89(3.39)&3.34(2.84)&2.92(2.45)\\
\hline
$\sigma_{\rm{P}}$&1.000&1.000&1.006&1.030\\
\hline
$\sigma_{\rm{T}}$&1.000&1.005&1.015&1.04\\
\hline
$\sigma_{\rm{PRF}}$&0.069&0.070&0.071&0.074\\
\hline
$\sigma_{\rm{TRF}}$&0.079&0.079&0.081&0.083\\
\hline
Gain&4.61$\pm$0.69&4.63$\pm$0.70&4.64$\pm$0.71&4.66$\pm$0.72\\
\hline
\end{tabular}
\caption{\label{tab:CFLossy} The influence of the lossy
compression with different lossy parameters to the cluster
characteristic. In row 1 effective range mapping shown. Entropy of
the samples are shown in row 2. Numbers in parenthesis represent
effective entropy od data sample, using different code table for
different sample position in the bunch. In row number 3 and 4
($\sigma_{\rm{P}}$ and $\sigma_{\rm{T}}$) the influence of the
lossy compression to the cluster space resolution in pad
respectively in time direction is shown. Row number 5 and 6 shows
the relative influence of compression to the shape of cluster in
time and pad directions. Gain row show the reconstructed ratio
between total deposited energy and numbers of contributing
electrons to the cluster. }
\end{center}
\end{table}

\begin{table*}[htb]
\begin{center}
\begin{tabular}{|l|c|c|c|c|}
\hline \textbf{} & \textbf{no} & \textbf{$K_{\rm{off}}$=1}
&\textbf{$K_{\rm{off}}$=1.5}&\textbf{$K_{\rm{off}}$=1}\\
\textbf{} & \textbf{} & \textbf{$K_{\rm{cor}}$=1}
&\textbf{$K_{cor}$=1.5}&\textbf{$K_{\rm{cor}}$=2}\\
\hline
$\sigma_{\phi}$[mrad]&1.399$\pm$0.030&1.378$\pm$0.030&1.406$\pm$0.030&
1.403$\pm$0.03\\
\hline
$\sigma_{\Theta}$[mrad]&0.997$\pm$0.018&0.992$\pm$0.018&1.002$\pm$0.018&
0.989$\pm$0.018\\
\hline
$\sigma_{p_{\rm{t}}}$[\%]&0.881$\pm$0.011&0.885$\pm$0.011&0.886$\pm$0.011&
0.905$\pm$0.011\\
\hline
$\sigma_{{\rm{d}}E/{\rm{d}}x}$[r.u]&2.96$\pm$0.11&2.98$\pm$0.11&3.06$\pm$0.11&
3.20$\pm$0.11\\
\hline
\end{tabular}
\caption{ \label{tab:TrackerLossy}The influence of the lossy
compression with different lossy parameters on the track
characteristics. }
\end{center}
\end{table*}

\subsection{Dynamic precision of the digitization}

In the following study, dynamic precision of sample quantization
was investigated. The quantization was chosen to correspond to the
sample deviation, modifying formula (\ref{eq:Sample}) to:
\begin{eqnarray}\
     \delta_{\rm{d}} = \sqrt{K_{\rm{off}}^2+K_{\rm{cor}}^2{\times}A}
        \label{eq:Sample2}
\end{eqnarray}
where $K_{\rm{off}}$ and $K_{\rm{cor}}$ factors were chosen as
free parameters. $K_{\rm{off}}$ is proportional to the electronic
noise and $K_{\rm{cor}}$ is given by statistics of the stochastic
processes and by correlations. Different combinations of these
factors were investigated.

In table \ref{tab:CFLossy}  the influence of different
quantization on the precision of the cluster characteristic
determination is shown.

The gain factor $G=A_{\rm{t}}/N_{\rm{el}}$ ($A_{\rm{t}}$ is the
total charge in cluster, $N_{\rm{el}}$ is the number of electrons
contributing to the cluster) measures the precision of the local
deposited energy determination. This factor is important for
d$E$d$x$ measurement and consequently for particle identification
(PID). The influence of the compression on the cluster position
determination varies between 0 to 4\%, depending on the
compression factor, as can be expected. The shape of the cluster
($\sigma_{\rm{PRF}}$ and $\sigma_{\rm{TRF}}$), important for
cluster quality determination, varies between 1 to 6\%.

In table \ref{tab:TrackerLossy} the influence of the compression
on the tracking is shown. Reported distortions in $p_{\rm{t}}$ and
angular resolution are slightly smaller than in the case of the
cluster position(0\% up to 3.5\%) . This is due to the other
stochastic processes which contribute to the track parameters e.g.
the multiple scattering. For high-momentum particles, where the
influence of multiple scattering is not so important, the expected
distortion will be determined by the cluster position distortions.

Reducing the number of the possible sample values, vector
quantization of bunches were also investigated. Additional
reduction factor of $\sim$6\% was achieved on top of the results
reported  in table \ref{tab:CFLossy}.

\section{Conlusions}

Several methods of lossless TPC data compression was investigated
(sec.~\ref{sec:Losslescomp}). The best one dimensional methods
provide compression factor down to 49.2\%.

A lossy compression approach for the data generated by the TPC
chamber in the ALICE experiment has been also investigated. The
main idea was to preserve, on the level intrinsic noise, the three
more important local quantities: the cluster position, the
deposited energy and the shape of the cluster. Keeping the
distortions of the local quantities at a reasonable level, the
impact on the most interesting physical quantities, particle
momenta and d$E$d$x$ is minimal.  This approach achieves
compression rates in the range from 35\% down to 30\%, depending
on the desired precision. In this study we have focused on methods
which are easy to implement in the frontend TPC electronics.

\bibliographystyle{unsrt}
\end{document}